\documentclass[onecolumn]{aa} 
\usepackage{graphicx}
\usepackage{natbib}
\bibpunct{(}{)}{;}{a}{}{,}
\usepackage{txfonts}
\begin{document}

   \title{Broad Ly$\alpha$ Emission from Supernova Remnants in Young Galaxies}
   \titlerunning{Broad Ly$\alpha$ from SNRs}
   \authorrunning{Heng \& Sunyaev}

   \author{Kevin Heng\inst{1,2} \& Rashid A. Sunyaev\inst{1,3}}

   \offprints{K. Heng,\\ e-mail: heng@ias.edu}

   \institute{Max Planck Institut f\"{u}r Astrophysik, 
              Karl-Schwarzschild-Stra$\beta$e 1, 85740 Garching, Germany
\and
              Max Planck Institut f\"{u}r extraterrestrische Physik,
              Giessenbachstra$\beta$e, 85478 Garching, Germany
\and
              Space Research Institute, Russian Academy of Sciences, 
              Profsoyuznaya 84/32, 117997 Moscow, Russia
             }

   \date{Received ...; ...}

  \abstract
   {Charge transfer (or exchange) reactions between hydrogen atoms and protons in collisionless shocks of supernova remnants (SNRs) are a natural way of producing broad Balmer, Lyman and other lines of hydrogen.}
   {We wish to quantify the importance of shock-induced, non-thermal hydrogen emission from SNRs in young galaxies.}
   {We present a method to estimate the luminosity of broad ($\sim 1000$ km s$^{-1}$) Ly$\alpha$, Ly$\beta$, Ly$\gamma$, H$\beta$ and P$\alpha$ lines, as well as the broad and narrow luminosities of the two-photon (2$\gamma$) continuum, from existing measurements of the H$\alpha$ flux.  We consider cases of $\beta=0.1$ and 1, where $\beta \equiv T_e/T_p$ is the ratio of electron to proton temperatures.  We examine a modest sample of 8 proximate, Balmer-dominated SNRs from our Galaxy and the Large Magellanic Cloud.  The expected broad Ly$\alpha$ luminosity per object is at most $\sim 10^{36}$ erg s$^{-1}$.  The 2$\gamma$ continuum luminosities are comparable to the broad H$\alpha$ and Ly$\alpha$ ones.  We restrict our analysis to homogenous and static media.}
   {Differences in the Ly$\alpha$/H$\alpha$ and Ly$\beta$/H$\alpha$
  luminosity ratios between the $\beta=0.1$ and 1 cases are factors
  $\sim 2$ for shock velocities $1000 \lesssim v_s \lesssim 4000$ km
  s$^{-1}$, thereby providing a direct and unique way to
  measure $\beta$.  In principle, broad, ``non-radiative'' Ly$\alpha$
  from SNRs in young galaxies can be directly observed in the optical
  range of wavelengths.  However, by taking into consideration the
  different rates between core collapse and thermonuclear
  supernovae, as well as the duration we expect to observe such Ly$\alpha$ emission from SNRs, we expect their
  contribution to the total Ly$\alpha$ luminosity from $z \sim 3$ to 5 galaxies to be negligibly small ($\sim 0.001 \%$),
  compared to the radiative shock mechanism described by Shull \& Silk (1979).  Though broad, non-thermal Ly$\alpha$ emission has never been observed, these photons are produced in SNRs and hence the non-radiative Ly$\alpha$ luminosity is a part of the intrinsic Ly$\alpha$ spectrum of young galaxies.}
   {}


   \maketitle

\section{Introduction}

Observations of galaxies at high redshifts have revealed a broad class
of Ly$\alpha$-emitting galaxies at $z \sim 3$ to 5 (e.g., Tapken et
al. 2007).  The Ly$\alpha$ emission from these objects is reaching us
as light in the visible spectral band, enabling their study using
large, ground-based optical telescopes, which in turn permits detailed spectroscopic studies of these galaxies.  Observations of quasars at $z \sim 6$ (e.g., Fan et al. 2006) have revealed heavy elemental abundances exceeding solar values.  We know that at least some of the galaxies at $z \sim 3$ to 5 have high abundances of heavy elements, facilitating the formation of dust.  In homogenous and static media, the dust particles impede the escape of Ly$\alpha$ emission from gas-rich galaxies, due to the small mean free paths of the photons, low temperatures of the gas and ultimately high probabilities of absorption.  In clumpy media, dust can enhance the escape of Ly$\alpha$ photons relative to the continuum (Neufeld 1991; Hansen \& Oh 2006).  Broadening of Ly$\alpha$ lines due to multiple scatterings is a slow process requiring a long diffusion time (though velocity fields in the interstellar medium may broaden the Ly$\alpha$ lines and reduce the diffusion time).  Hence, there is special interest in the physical processes that are able to naturally produce extremely broad wings in Ly$\alpha$ lines, which may permit the photons to leave the host galaxy without requiring many scatterings (but see \S\ref{sect:discussion}).

Among obvious mechanisms is the one at work in the unique massive
binary SS433 (for a recent review, see Fabrika [2004]), with strongly blue- and redshifted
H$\alpha$ and H$\beta$ lines, due to cooling and recombination of
hydrogen in the baryon-dominated, precessing jet moving with velocity
$\sim 0.26 c$.  Such objects are very rare --- SS433 is the only such
example in our Galaxy.  More well-known Galactic sources of H$\alpha$
emission with broad line wings are the supernova remnants (SNRs) of
Type Ia, emitting due to charge transfer (or ``charge exchange'')
reactions between hydrogen atoms and protons in the blast wave
penetrating the low-density ($\sim 1$ cm$^{-3}$), ambient gas.  The
widths of the H$\alpha$ lines correspond to Doppler broadening with
velocities up to $\sim 5000$ km s$^{-1}$.  The same process should
produce not only H$\alpha$ emission, but photons in the Lyman series
of hydrogen as well.  Recently, some of these SNRs were observed in Ly$\beta$
using the {\it FUSE} spacecraft (Korreck et al. 2004; Ghavamian et al. 2007, hereafter G07).

Knowledge of the cross sections of charge transfers to excited levels
and excitation of the fast-moving hydrogen atoms permit us to find
simple formulae relating the luminosities of SNRs in the broad
H$\alpha$ and Ly$\alpha$ lines.  The Ly$\alpha$ line should have a similar spectral distribution to the observed H$\alpha$ one in the broad wings, because the optical depth of the SNR for broad photons is negligibly small and the optical depth for coherent scattering (in the distant Lorentzian wings) in interstellar gas is low.

We compile the existing data for core collapse and thermonuclear SNRs,
including SNR 1987A (where the reverse shock is bright in the broad
H$\alpha$ line), and present their theoretically expected, broad
Ly$\alpha$ and Ly$\beta$ luminosities.  For two objects, we present
their expected broad Ly$\gamma$, H$\beta$ and P$\alpha$ luminosities.
Taking into account the supernova (SN) rates, the luminosities of the
SNRs in H$\alpha$ and the duration of their active phase (for the
charge transfer mechanism described), we find that --- even without
discussing the cosmological evolution of the SN rates --- the expected
broad Ly$\alpha$ is several orders of magnitude lower than the
estimate of Shull \& Silk (1979), who treated fully radiative SNRs with low
metallicities and velocities (20 to 120 km s$^{-1}$).  We come to the conclusion that the contribution of both core collapse and thermonuclear SNRs to the Ly$\alpha$ luminosity of young galaxies is negligibly small.  In \S\ref{sect:obs}, we gather a modest sample of 8 Galactic and Large Magellanic Cloud (LMC) remnants, and use them as a template for estimating the expected Ly$\alpha$, Ly$\beta$, Ly$\gamma$, H$\beta$ and P$\alpha$ production.  In \S\ref{sect:ratios}, we compute the Ly$\alpha$/H$\alpha$, Ly$\alpha$/Ly$\beta$, Ly$\beta$/H$\alpha$, Ly$\gamma$/H$\alpha$, H$\beta$/H$\alpha$ and P$\alpha$/H$\alpha$ luminosity ratios.  We present our results in \S\ref{sect:results} and discuss their implications in \S\ref{sect:discussion}.

\section{Galactic \& LMC Remnants}
\label{sect:obs}

SNRs are the result of the interaction of SN ejecta with ambient matter.  The nature of the interaction can be approximately categorized into several stages (Truelove \& McKee 1999, hereafter TM99; and references therein): the ejecta-dominated (ED) or freely-streaming stage; the Sedov-Taylor (ST) or self-similar stage; the pressure-driven snowplow (PDS) stage; and a possible, momentum-conserving snowplow stage (Cioffi, McKee \& Bertschinger 1988).  Many of the well-studied, young SNRs like Kepler, Tycho and SN 1006 are intermediate between the ED and ST stages; this has been corroborated by the numerical studies of TM99, who showed that there is no sharp transition between the two stages.  The transition from the ED to ST stage occurs on a timescale $t_{\rm{SD}} \sim t_{\rm{ch}}$; the characteristic timescale is
\begin{equation}
t_{\rm{ch}} = t_{\rm{ch},0} ~m^{5/6}_{\rm{ej}} E^{-1/2}_{51}
n^{-1/3}_0,
\label{eq:chartime}
\end{equation}
where $M_{\rm{ej}} = m_{\rm{ej}} M_{\sun}$ is the mass of the ejecta, $E
= E_{51} 10^{51}$ erg is the energy of the supernova explosion, and
$n_0$ is the density of the ambient medium (in cm$^{-3}$).  The
coefficient in equation (\ref{eq:chartime}) is $t_{\rm{ch},0}= 423$
yrs (TM99).  If one makes the argument that $M_{\rm{ej}}$ (with
density $\rho = 1.4 m_{\rm{H}} n_0$) of mass is swept up in a time
$t_{\rm{ch}}$, one instead gets $t_{\rm{ch},0}=186$ yrs.  The PDS stage occurs at
\begin{equation}
t_{\rm{PDS}} \approx 30 t_{\rm{ch}} ~m^{-5/6}_{\rm{ej}} E^{5/7}_{51} n^{-5/21}_0 \zeta_m^{-5/14}
\end{equation}
after the explosion (Cioffi, McKee \& Bertschinger 1988; TM99), where $\zeta_m$ is a dimensionless metallicity correction factor.  More precise estimates for $t_{\rm{SD}}$ and $t_{\rm{PDS}}$ are dependent upon the spatial density distributions of both the ejecta and the ambient matter.

In the ED and SD stages, the emission from some SNRs is
``non-radiative'', meaning the timescale for thermal, radiative losses from the
interacting gases is much longer than $t_{\rm{ch}}$.  When the blast
wave of the SNR slams into ambient gas consisting predominantly of
hydrogen atoms, it emits in Balmer and Lyman lines consisting of a
broad ($\sim 1000$ km s$^{-1}$) and a narrow ($\sim 10$ km s$^{-1}$)
component (Chevalier \& Raymond 1978; Bychkov \& Lebedev 1979; Chevalier, Kirshner \& Raymond 1980; Heng
\& McCray 2007, hereafter HM07; Heng et al. 2007, hereafter H07; G07; and references therein).  These
objects are known as ``Balmer-dominated'' SNRs.  Positive detections
of the line components are so far only from Galactic and LMC SNRs.
Even though narrow Ly$\alpha$ emission is produced, it is not seen due
to interstellar absorption; broad Ly$\alpha$ should be observed\footnote{G07 recorded {\it FUSE} spectra only in the 905 --- 1100 and 987 --- 1180 \AA bands.}.    Non-thermal H$\alpha$ and Ly$\alpha$ emission has not been observed in studies of local starburst galaxies (e.g., Kunth et al. 2003).

The narrow Balmer and Lyman lines are produced when the fast-moving
ejecta directly excite stationary hydrogen atoms in the
surrounding material.  The broad lines are produced when the
post-shock protons and atoms engage in charge transfer reactions,
creating a population of post-shock atoms in broad velocity
distributions known as ``broad neutrals'' (HM07; H07).  In the frame
of the observer, these broad neutrals move at a velocity $v_{\rm{B}}
\lesssim 3v_s/4$, where $v_s$ is the shock velocity (of the blast
wave).  For $v_s \gtrsim 500$ km s$^{-1}$, the broad neutrals can
produce Ly$\alpha$ that is blue- or redshifted out of resonance with
the stationary atoms, hence providing an escape route for the photons.
The ratio of broad to narrow H$\alpha$ (and Ly$\alpha$) emission is a
function of the shock velocity (HM07; H07); it also depends on factors
like the pre-shock neutral density and the degree to which the
temperatures of the electrons and ions are equilibrated.  The
contribution from the broad H$\alpha$ line dominates when the shock
velocity is $\lesssim 3000$ km s$^{-1}$ and when the narrow
H$\alpha$ line assumes Case A conditions (HM07).  Existing
observations of H$\alpha$ and Ly$\beta$ emission from 8
Balmer-dominated SNRs are catalogued in Table \ref{table:obs}.  At
least 5 of these SNRs are believed to have resulted from Type Ia
explosions.  Only SNR 1987A has a clear core collapse origin; it is also the youngest SNR in the sample.

To convert H$\alpha$ line fluxes to broad Ly$\alpha$ luminosities, we use
\begin{equation}
L_{\rm{Ly}\alpha} = 4 \pi d^2 F_{\rm{H}\alpha} ~\frac{\Re_{bn}}{1+\Re_{bn}} ~\Gamma_{\rm{Ly}\alpha/\rm{H}\alpha},
\label{eq:lum}
\end{equation}
where $d$ is the distance to the SNR and $\Re_{bn} \sim 1$ is the observed ratio of
broad to narrow H$\alpha$ emission.  The quantity
$\Gamma_{\rm{Ly}\alpha/\rm{H}\alpha}$ is the ratio of Ly$\alpha$ to
H$\alpha$ luminosities (see \S\ref{sect:ratios}).  For SNRs in the
LMC, we adopt $d=50$ kpc.  In the case of the LMC remnant 0509---67.5,
$\Re_{bn}$ is unavailable, so we quote an upper limit.  If several
values for the H$\alpha$ flux are given, we simply choose the
brightest one (e.g., different emission knots of Kepler's SNR).  For
SNR 1987A, we take the observed value of $\Re_{bn} \sim 1$ (Heng et
al. 2006), as opposed to the theoretically calculated one ($\sim 0.1$;
HM07); we use the measured H$\alpha$ flux to obtain an estimate for
the broad Ly$\alpha$ luminosity, as the measured Ly$\alpha$ flux is
subjected to resonant scattering.  The Cygnus Loop is excluded from
our sample due to its low shock velocity of $\sim 250$ km s$^{-1}$.
In using equation (\ref{eq:lum}), we note that the measured H$\alpha$
and Ly$\beta$ fluxes are mostly from limb-brightened portions of the
SNRs.  Assuming spherical remnants, the intensities from these parts
are brightened by factors $\sim (R/l_a)^{1/2}$ (Chevalier \& Raymond
1978; H07), where $R \sim 1$ pc is the typical radius of the SNR and
$l_a \sim 10^{15}$ cm is the length scale for atomic interactions
(assuming density $\sim 1$ cm$^{-3}$ and velocity $\gtrsim 1000$ km
s$^{-1}$).  Hence, the luminosity inferred might be over-estimated by
a factor $\sim 50$.

\section{THE Ly$\alpha$/Ly$\beta$ \& Ly$\alpha$/H$\alpha$ Ratios}
\label{sect:ratios}

\begin{figure}
  \resizebox{\hsize}{!}{\includegraphics{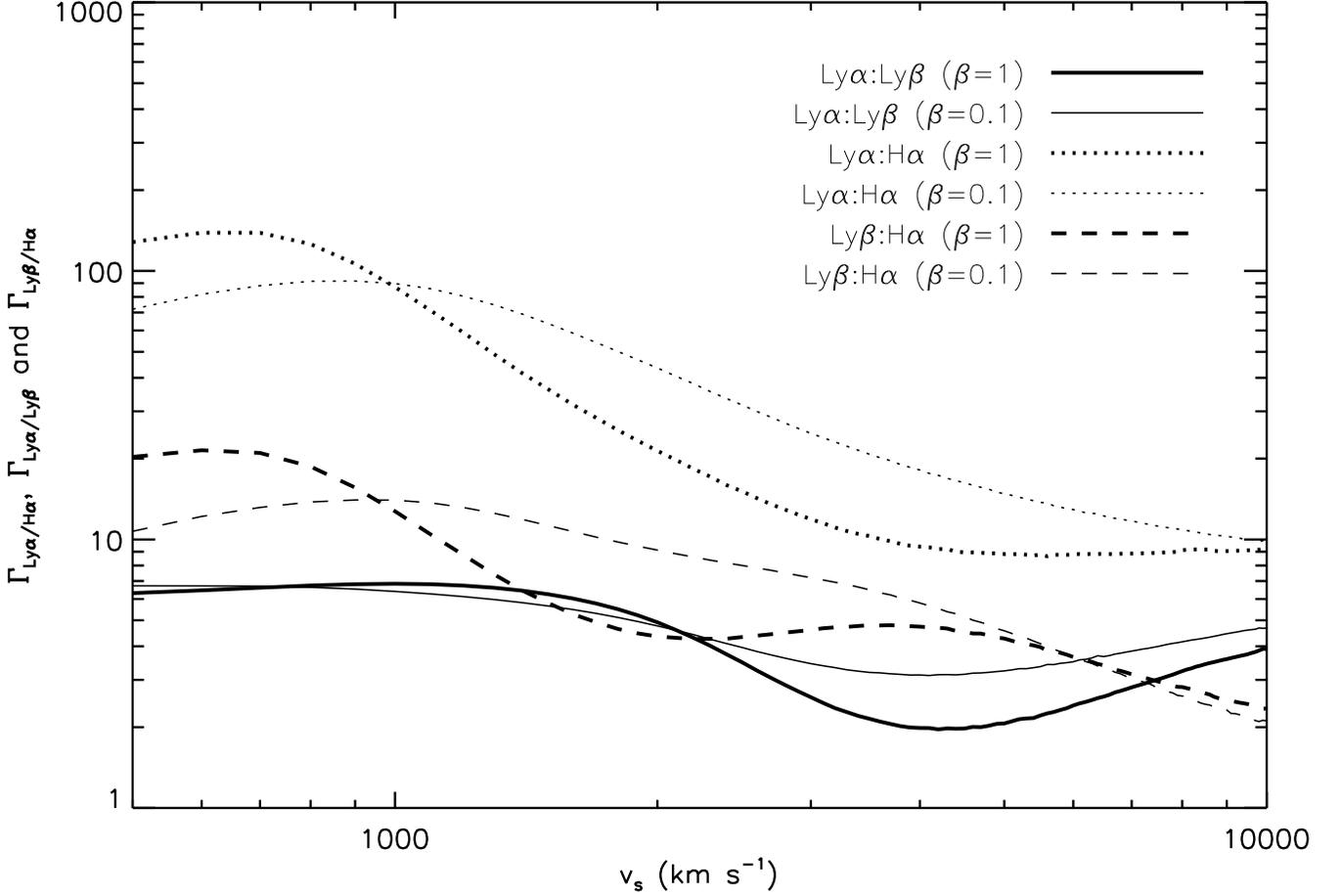}}
  \caption{Luminosity ratios of Ly$\alpha$ to H$\alpha$, Ly$\alpha$ to Ly$\beta$, and Ly$\beta$ to H$\alpha$, denoted by $\Gamma_{\rm{Ly}\alpha/\rm{H}\alpha}$, $\Gamma_{\rm{Ly}\alpha/\rm{Ly}\beta}$ and $\Gamma_{\rm{Ly}\beta/\rm{H}\alpha}$, respectively, as a function of the shock velocity, $v_s$.}
  \label{fig:ratios}
\end{figure}

The ratio of Ly$\alpha$ to H$\alpha$ luminosities (Fig. \ref{fig:ratios}) is computed using the methods developed by HM07:
\begin{equation}
\Gamma_{\rm{Ly}\alpha/\rm{H}\alpha}\left(nl,n^\prime l^\prime\right) =  \frac{\epsilon \left( R_{E,nl} + R_{T^*,nl} \right) + R_{T^*_0,nl}}{\epsilon \left( R_{E,n^\prime l^\prime} + R_{T^*,n^\prime l^\prime} \right) + R_{T^*_0,n^\prime l^\prime}},
\label{eq:lumratio}
\end{equation}
where $\epsilon = P_{T_0}/P_I$.  The quantity $P_{T_0}$ is the
probability for pre-shock atoms (found in a beam, i.e., at one
velocity) to engage in charge transfer reactions with ions (thereby
creating broad neutrals), while
$P_I$ is the probability for the broad neutrals to be ionized by both
electrons and ions.  Physically, broad H$\alpha$ emission is produced
in two ways: charge transfer of the pre-shock atoms to excited states
of broad neutrals (with a rate coefficient, in cm$^3$ s$^{-1}$, of
$R_{T^*_0,nl}$); creation of broad neutrals in the ground state,
followed by excitation ($R_{E,nl}$) and/or charge transfers between
them and ions to excited states ($R_{T^*,nl}$).   Hence, $\epsilon$ is
a measure of how efficient the first contribution is relative to the
second one.  At low shock velocities ($v_s \lesssim 1000$ km
s$^{-1}$), $\epsilon \gtrsim 3$ --- charge transfer to the ground
state is the dominant process, and it is efficient to create broad
neutrals that subsequently get excited.  We emphasize that equation
(\ref{eq:lumratio}) is only valid in the case of optically-thin
plasmas.

For Ly$\alpha$, we consider charge transfers (with protons) and
excitations (by electrons and protons) to the sub-levels $2p$, $3s$ and
$3d$.  For H$\alpha$, we consider the same processes, but for the
sub-levels $3s$, $3p$ and $3d$.  Hence, we compute
$\Gamma_{\rm{Ly}\alpha/\rm{H}\alpha}(nl,n^\prime l^\prime)$ for $nl=2p+3s+3d$ and
$n^\prime l^\prime=3s+3d+ B_{3p,2s} 3p$, where the factor of $B_{3p,2s}=0.1183$ is the
fraction of radiative decays from $3p$ that result in H$\alpha$, with
the remainder going to Ly$\beta$.  For
$\Gamma_{\rm{Ly}\alpha/\rm{Ly}\beta}(nl,n^\prime l^\prime)$, we
consider instead $n^\prime l^\prime=(1-B_{3p,2s})3p$.  Cascade
contributions from higher levels are $\lesssim 5\%$ effects.  For
example, contributions to H$\alpha$ from $n=4$ are at most $\sim
(3/4)^3 B_{4s,3p} B_{3p,2s} \approx 2\%$; other contributions from
$4p$, $4d$ and $4f$ are at the $\lesssim 1\%$ level.

\begin{figure}
  \resizebox{\hsize}{!}{\includegraphics{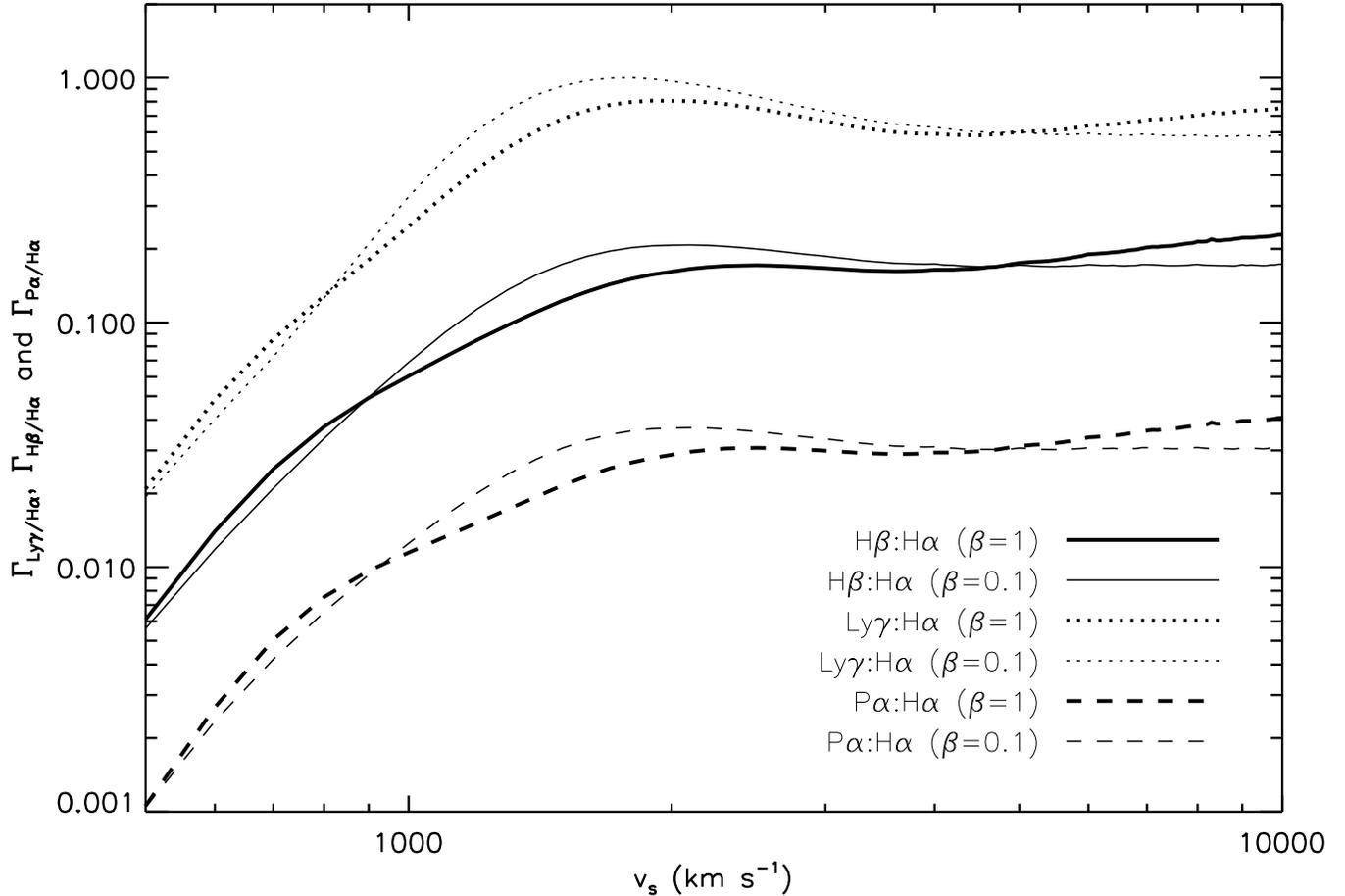}}
  \caption{Luminosity ratios of Ly$\gamma$ to H$\alpha$, H$\beta$ to
H$\alpha$, and P$\alpha$ to H$\alpha$, denoted by $\Gamma_{\rm{Ly}\gamma/\rm{H}\alpha}$, $\Gamma_{\rm{H}\beta/\rm{H}\alpha}$ and $\Gamma_{\rm{P}\alpha/\rm{H}\alpha}$, respectively, as a function of the shock velocity, $v_s$.  Only charge transfers to excited states are considered in these ratios, so they should be used with caution; we use only the luminosity ratios for $v_s \gtrsim 5000$ km s$^{-1}$.}
  \label{fig:ratios2}
\end{figure}

One can calculate the luminosity ratios for Ly$\gamma$/H$\alpha$,
H$\beta$/H$\alpha$ and P$\alpha$/H$\alpha$ as well.  However, the
cross sections for impact excitation of hydrogen atoms by protons
to the sub-levels $4s$, $4p$, $4d$ and $4f$ are unavailable at the time of
writing\footnote{We note that Mart\'{i}n (1999) has computed the cross sections for impact excitation of hydrogen atoms by protons to the sub-levels $4s$, $4p$, $4d$ and $4f$, but only for energies of 30 to 200 keV.}.  The cross sections for charge transfers to these excited
states, however, are available.  At $v_s \gtrsim 5000$ km s$^{-1}$,
$\epsilon \lesssim 0.5$, and we may obtain luminosity ratios for
Ly$\gamma$/H$\alpha$, H$\beta$/H$\alpha$ and P$\alpha$/H$\alpha$ to
within a factor of 2 (Fig. \ref{fig:ratios2}).   A list of the
relevant radiative decay fractions, $B_{nl,n^\prime l^\prime}$, is given in Table
\ref{table:einstein} (see Appendix \ref{append:einstein} for details).  In principle, if the charge transfer and excitation cross sections are known to higher levels in the relevant velocity range, one can calculate the luminosity ratios for other lines in the Balmer, Lyman, Paschen and other series of hydrogen.

We use the atomic cross sections of Balan\c{c}a, Lin \& Feautrier
(1998), Barnett et al. (1990), Belk\'{i}c, Gayet \& Salin (1992),
Harel, Jouin \& Pons (1998) and Janev \& Smith (1993), as well as
those found in the {\it NIST Electron-Impact Cross Section Database}.
Details concerning the cross sections are given in Appendix
\ref{append:atomic}, where we provide fitting functions to them.  We consider a pure hydrogen gas and include charge transfer, excitation and ionization events between hydrogen atoms, electrons and protons.  We employ the thin shock approximation, such that the relative velocity between atoms and ions is $3v_s/4$; this has been shown by H07 to be an excellent approximation.  At the shock velocities considered, $500 \lesssim v_s \lesssim 10,000$ km s$^{-1}$, the significance of impact excitation by electrons is comparable to that by protons and cannot be neglected.  We do not consider broad emission from within the shock front (see Appendix \ref{append:within}).

\section{Results}
\label{sect:results}

The luminosities ratios $\Gamma_{\rm{Ly}\alpha/\rm{H}\alpha}$,
$\Gamma_{\rm{Ly}\alpha/\rm{Ly}\beta}$ and
$\Gamma_{\rm{Ly}\beta/\rm{H}\alpha}$ are shown in
Fig. \ref{fig:ratios}.  In the shock velocity range $1000 \lesssim v_s
\lesssim 4000$ km s$^{-1}$, the differences in
$\Gamma_{\rm{Ly}\alpha/\rm{H}\alpha}$ and
$\Gamma_{\rm{Ly}\beta/\rm{H}\alpha}$ between the $\beta=0.1$ and 1
cases are factors $\sim 2$, and are due to the sensitivity to
temperature of impact
excitation and ionization of hydrogen atoms by electrons.  This may present a direct and unique
opportunity to measure $\beta$.  We emphasize that our calculations
are only valid for the broad lines; the narrow lines have optical
depths $0 \le \tau \le 1$ and Lyman line trapping is a non-negligible
effect (Ghavamian et al. 2001, 2002).  For example, narrow Ly$\beta$
photons may be converted into narrow H$\alpha$ photons and two-photon
(2$\gamma$) continuum.  In addition, narrow Ly$\alpha$ cannot
propagate easily through the interstellar gas.

\begin{figure}
  \resizebox{\hsize}{!}{\includegraphics{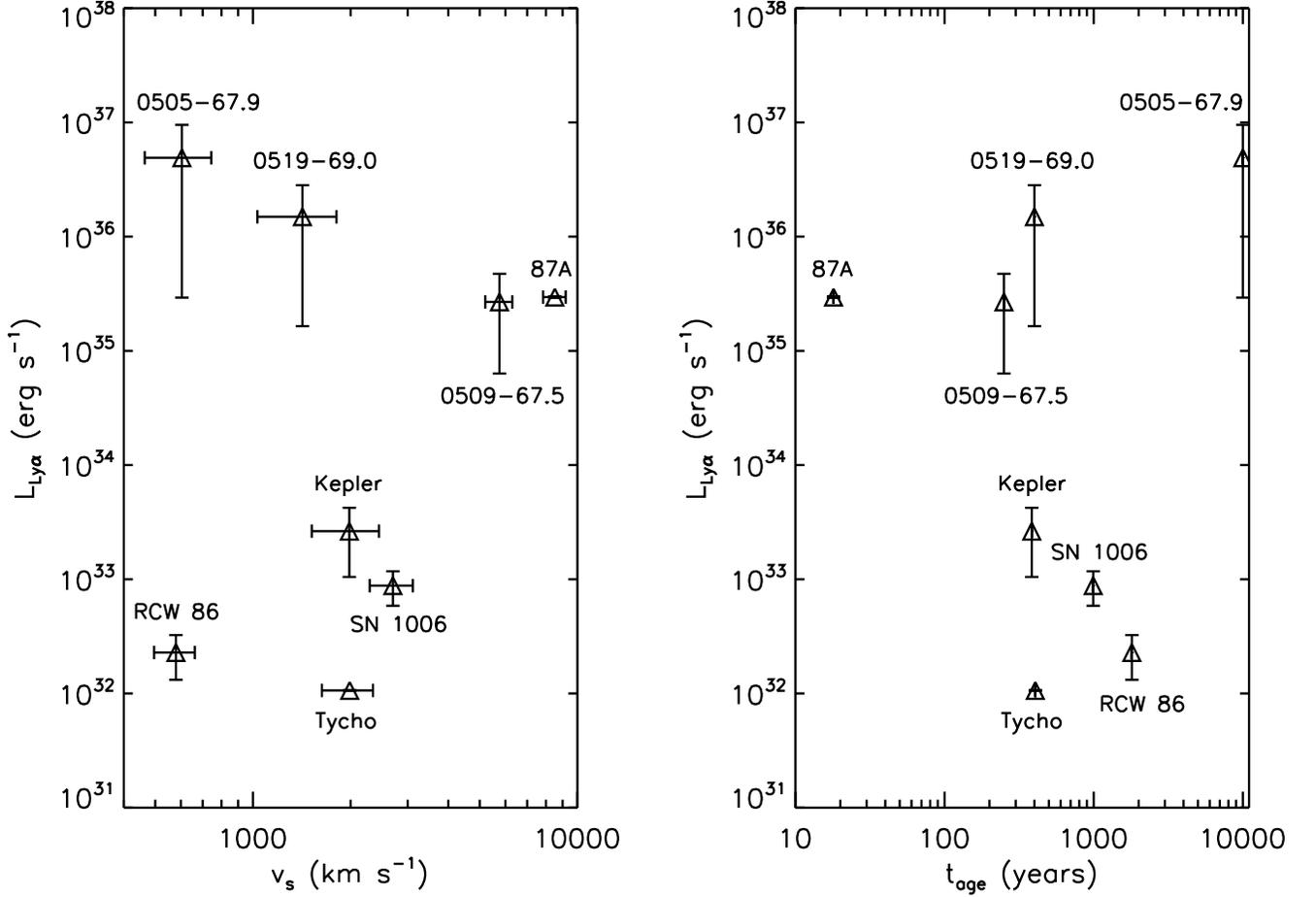}}
  \caption{Left: Expected Ly$\alpha$ luminosities, $L_{\rm{Ly}\alpha}$, from the SNR sample as a function of the shock velocity, $v_s$.  Right: $L_{\rm{Ly}\alpha}$ plotted versus the age of the SNR.}
  \label{fig:lum}
\end{figure}

We use the data in Table \ref{table:obs} to compute the expected
luminosity of Ly$\alpha$, $L_{\rm{Ly}\alpha}$ (Fig. \ref{fig:lum}).
In estimating a range for $L_{\rm{Ly}\alpha}$, we only consider the
observational error bars in $F_{\rm{H}\alpha}$ (if available) and
allow for a generous range in temperature equilibration between
electrons and protons, $0.1 \le \beta \le 1$, where
$\beta = T_e/T_p$.  Hence, the displayed error bars for
$L_{\rm{Ly}\alpha}$ are not formal ones.  We are aware of the recent
work by Ghavamian, Laming \& Rakowski (2007), who showed that there is
an empirical correlation between $\beta$ and $v_s$ --- namely,
$\beta=1$ for $v_s \lesssim 400$ km s$^{-1}$ and $\beta \propto
v^{-2}_s$ for $v_s \gtrsim 400$ km s$^{-1}$.  For the LMC remnants
detected in Ly$\beta$ by G07, we compute the range in
$L_{\rm{Ly}\alpha}$ by considering both the H$\alpha$ and Ly$\beta$
fluxes.  We note that the computed $(2.96 \pm 0.05) \times 10^{35}$ erg s$^{-1}$
value for broad Ly$\alpha$ in SNR 1987A is comparable to the $\sim
10^{36}$ erg s$^{-1}$ figure predicted by Michael et al. (2003).  Note
that the condition $L_{\rm{L}\beta}/L_{\rm{H}\alpha} \ge
\lambda_{\rm{H}\alpha}/\lambda_{\rm{L}\beta} \approx 6.4$ is not true
in general.  This is because the cross section for charge transfers to
the level $3p$ falls below that to $3s$ at a relative velocity $\sim
2000$ km s$^{-1}$ (Fig. \ref{fig:pct3}).

In Table \ref{table:predict}, we make some predictions for the Ly$\beta$, Ly$\gamma$, H$\beta$ and P$\alpha$ luminosities.  It is puzzling that the theoretically expected Ly$\beta$ luminosities are about 10 to 20 times higher than those inferred from the observations of G07.  In other words, the observed H$\alpha$ fluxes in the LMC SNRs are comparable to the observed Ly$\beta$ ones.  We are not certain why this is the case, but we note that Ly$\beta$ is more susceptible to absorption by interstellar dust than H$\alpha$, and we suspect this effect to play at least some part in the discrepancy.  Moreover, the H$\alpha$ and Ly$\beta$ observations were taken at different epochs (Tuohy et al. 1982 versus Ghavamian et al. 2007).  As described in \S\ref{sect:ratios}, we are only able to provide rough predictions for Ly$\gamma$, H$\beta$ and P$\alpha$, and only in the cases of 0509---67.5 and SNR 1987A, as these SNRs have shock velocities $\gtrsim 5000$ km s$^{-1}$.

\begin{figure}
  \resizebox{\hsize}{!}{\includegraphics{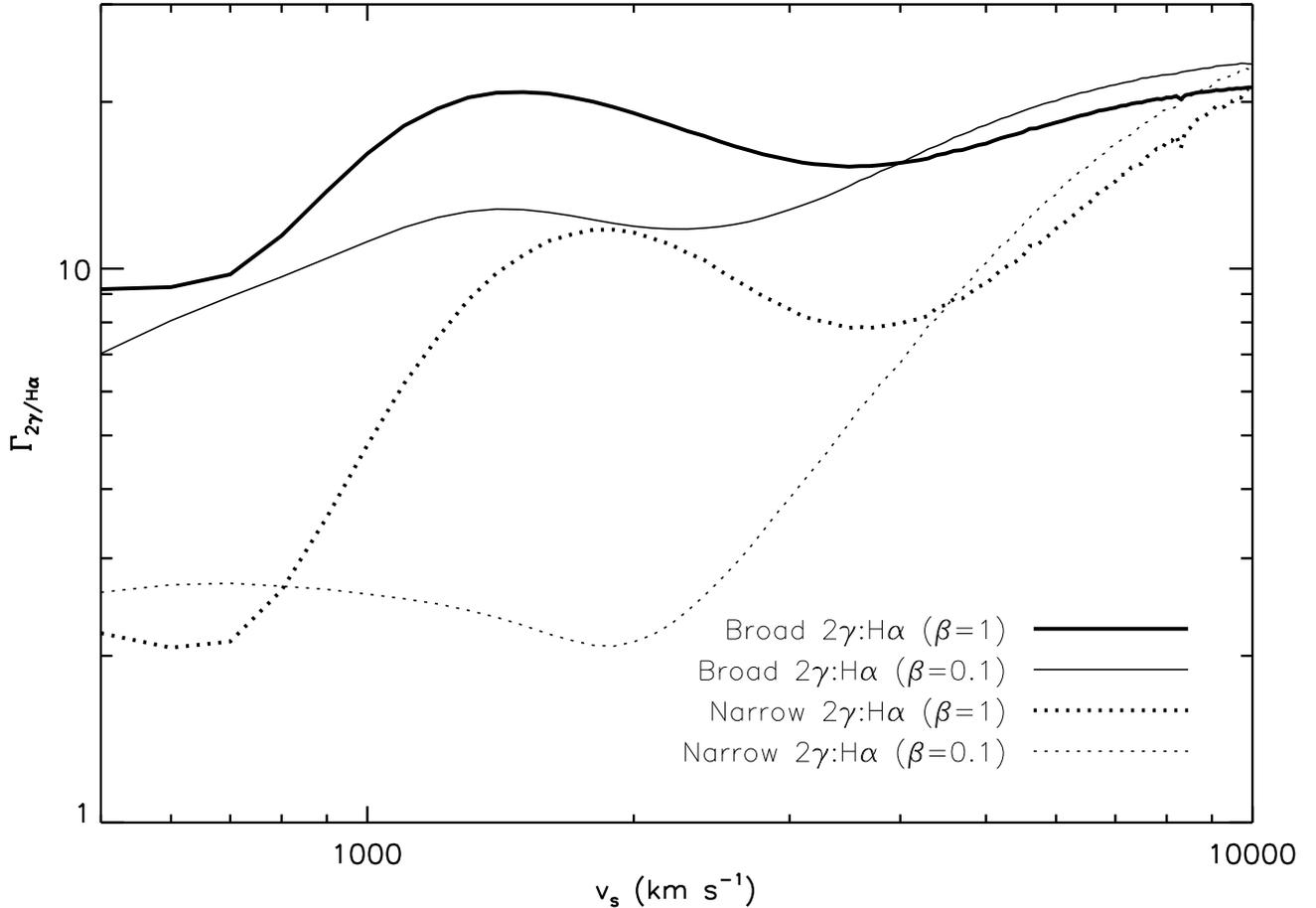}}
  \caption{Luminosity ratios of the 2$\gamma$ continuum to
  H$\alpha$, $\Gamma_{2\gamma/\rm{H}\alpha}$.  Only charge transfers
  and excitations to the $2s$ level are considered (see text).}
  \label{fig:ratios3}
\end{figure}

We can make some estimates for the expected 2$\gamma$ continuum as
well, which is produced in the $2s \rightarrow 1s$ transition.  In the
case of an optically-thin plasma, the $2s \rightarrow 2p$ transition
is negligible as collisions are unimportant.  In Table
\ref{table:predict}, we make conservative predictions for the
2$\gamma$ continuum luminosity from both broad and narrow atoms, but consider only charge transfers
and excitations to the $2s$ level.  Additional contributions from
$n=3$ range from $\sim (2/3)^3 B_{3p,2s} \approx 4\%$ (Case A) to
$\sim (2/3)^3 \approx 30\%$ (Case B); those from $n=4$ are $\ll 1\%$.
We only wish to make the point that $L_{2\gamma}$ is comparable to
$L_{\rm{Ly}\alpha}$ and  $L_{\rm{H}\alpha}$, and thus the 2$\gamma$
transitions are a potentially observable source of continuum.  In the
case of Galactic and LMC SNRs, the {\it Galaxy Evolution Explorer (GALEX)}
is in principle able to measure the low-frequency wing of 2$\gamma$
decay, using its 135---175 and 175---280 nm channels.  By comparing
H$\alpha$ and 2$\gamma$ emission, it will be possible to directly
estimate emission from the SNR shock due to broad Ly$\alpha$ (and the
contribution of narrow Ly$\alpha$ that cannot reach us).  This is an
additional, unique source of information on the detailed physical
processes in shocks.

Several sources of uncertainty can affect the predicted luminosities.  These include uncertainties in the age of the
SNR, $t_{\rm{age}}$, the distance to it, $d$, the measured {\it
non-radiative} component of the H$\alpha$ flux, the temperature
equilibration between electrons and ions, and the atomic cross
sections used.   Uncertainties in the cross sections are typically
$\sim$ 10\%.  For charge transfer to excited states, the uncertainty
can be as much as 30\% (R.K. Janev 2007, private communication).  The
predicted luminosities have not been corrected for reddening by dust.

\section{Discussion}
\label{sect:discussion}

SNR 1987A is a unique example of a Balmer-dominated SNR.  By virtue of
adiabatic expansion cooling, the SN ejecta comprises mostly neutral
hydrogen; it rushes out at velocities $\gtrsim 12,000$ km s$^{-1}$
(Michael et al. 2003; Heng et al. 2006).  The non-radiative H$\alpha$
and Ly$\alpha$ result from the interaction of the ejecta with the {\it
reverse shock} and not the blast wave (Heng 2007).  As SNR 1987A has a
Type II origin, it is possible to produce Balmer and Lyman lines
via this mechanism; this is obviously not possible with Type Ia's.
Smith et al. (2005) have predicted that the H$\alpha$ and Ly$\alpha$
emission from the reverse shock of SNR 1987A is shortlived ($\sim$
2012 to 2014) and will be extinguished by the increasing flux of
extreme ultraviolet (EUV) and X-ray photons traveling into the
pre-shock region and ionizing the atoms --- pre-ionization.  This is marginal evidence that broad Ly$\alpha$ from SNRs of a core collapse origin will be short-lived, i.e., $\lesssim 100$ years.  In general, for this scenario to work, some interaction of the blast wave with the ambient material is needed, but if it is too strong the pre-shock gas becomes ionized (R. Chevalier 2007, private communication).

To further investigate the viability of the short-lived, non-radiative Ly$\alpha$ hypothesis, we examine the sample of optically identified SNRs by Matonick \& Fesen (1997), who studied an ensemble of 12 SNR samples from different galaxies, including the Small Magellanic Cloud (SMC), LMC, M31 and M33, with distances up to 7 Mpc.  In galaxies like NGC 2403, M81 and M101, the SNRs are associated with star-forming regions and most of them probably have a Type Ib/c origin.  In most cases, the measured H$\alpha$ flux is $\sim 10^{-15}$ erg cm$^{-2}$ s$^{-1}$ and the inferred luminosity is $\sim 10^{36}$ erg s$^{-1}$.  Since Matonick \& Fesen (1997) did not provide H$\alpha$ line profiles, it is impossible to estimate the proportion of the H$\alpha$ emission that is non-radiative.  Furthermore, their selection criterion is based on picking out objects with [S~{\sc ii}]/H$\alpha \ge 0.45$, which will not detect SNRs with predominantly non-radiative H$\alpha$ emission. 

Shull \& Silk (1979) computed the temporally-averaged Ly$\alpha$ luminosity
from radiative shocks of a population of Type II SNRs, assuming low
metallicities, to be
\begin{equation}
L_{\rm{SS79}} = 3 \times 10^{43} \mbox{ erg s}^{-1} E^{3/4}_{51} n^{-1/2}_0 \dot{N}_{\rm{SN}}, 
\end{equation}
where $\dot{N}_{\rm{SN}}$ is the number of supernovae (SNe) a year.  They considered SNRs in both the ST and the PDS stages, and $v_s = 20$ to 120 km s$^{-1}$.  Charlot \& Fall (1993) remark that the numerical coefficient in the preceding equation is about 40\% lower if one assumes solar metallicity.

A very conservative upper limit on the broad Ly$\alpha$ from the
Matonick \& Fesen (1997) samples can be obtained if one generously
allows for all of the H$\alpha$ to be broad, for the shock velocities
to be low ($\sim 500$ km s$^{-1}$) such that
$\Gamma_{\rm{Ly}\alpha/\rm{H}\alpha} \sim 100$, and for the
non-radiative emission to last $\sim 10^4$ years.  Even in this very
unlikely scenario, $L_{\rm{Ly}\alpha} \sim 10^{42}$ erg s$^{-1}$ is
only about $0.1 L_{\rm{SS79}}$.  Hence, our charge transfer mechanism
is not energetically competitive.  There is the possibility a SNR can produce both radiative and
non-radiative components of H$\alpha$.  Well-known examples are Kepler
(Fesen et al. 1989; Blair, Long \& Vancura 1991) and RCW 86 (Long \&
Blair 1990; Smith 1997).  There is also the possibility that the
non-radiative emission from the SNR is inhibited.  For example, Foster (2005) observed and studied the Galactic SNR 3C 434.1 ($t_{\rm{age}} \approx 25,000$ yr; $d = 4.5 \pm 0.9$ kpc; possible Type Ib/c), which formed inside the eastern portion of a pre-existing stellar-wind bubble of interior density $\sim 0.1$ cm$^{-3}$.  Strong H$\alpha$ emission ($6.1 \pm 0.4 \times 10^{36}$ erg s$^{-1}$) is measured from the eastern side; it is believed to be from a radiative shock.  Being farther away from the western wall of the bubble, the shock on the western side is essentially still in free expansion and produces no measurable, non-radiative H$\alpha$.

Our SNR sample and the considerations of SNR 1987A lead us to believe that if the short-lived emission contribution from Type Ib/c and Type II SNRs in young galaxies exists, it has a luminosity of
\begin{equation}
L_{\rm{Ly}\alpha,\rm{CC}} \sim 10^{38} \mbox{ erg s}^{-1} ~t_{\rm{emit},2} \dot{N}_{\rm{SN}},
\end{equation}
where $t_{\rm{emit}} = t_{\rm{emit},2} 100$ years is the length of
time we expect core collapse SNRs to produce shock-induced Ly$\alpha$
emission.  On the other hand, thermonuclear SNRs are expected to have  $t_{\rm{emit}} = t_{\rm{emit},4} 10^4$ years $\sim t_{\rm{PDS}}$.  However, they are also believed to be much scarcer at high redshifts.  For example, Dahlen et al. (2004) estimate that only 5\% to 7\% of available progenitors explode as Type Ia SNRs.  Therefore, the expected luminosity is
\begin{equation}
L_{\rm{Ly}\alpha,\rm{Ia}} \sim 10^{38} \mbox{ erg s}^{-1} ~t_{\rm{emit},4} \dot{N}_{\rm{SN},-2},
\end{equation}
where $\dot{N}_{\rm{SN},-2}$ is the number of SN per year in units of
0.01.  We conclude that for both core collapse and thermonuclear SNRs,
the expected luminosity from broad Ly$\alpha$ is only a $\sim 0.001\%$
effect, compared to the mechanism of Shull \& Silk (1979).  Ly$\alpha$
line luminosities from $z \sim 3$ to 5 galaxies have been
observationally determined to be $\sim 10^{42}$ to $10^{43}$ erg
s$^{-1}$ (e.g., Saito et al. 2007), in general agreement with theoretical expectations.  In addition, the lifetime of an emitting atom is approximately the length of time corresponding to one atomic length scale, and is only $t_{\rm mfp} \sim l_a/v_s \sim 10^7 n^{-1}_0 v^{-1}_{s,8}$ s, where $v_{s,8} = v_s/1000$ km s$^{-1}$ (H07).

We have restricted our analysis to homogeneous and static media.  Though broad, non-thermal Ly$\alpha$ emission has never been observed, these photons {\it are} produced in SNRs and hence the non-radiative Ly$\alpha$ luminosity is a part of the intrinsic Ly$\alpha$ spectrum of young galaxies.  The optical depth for a broad photon in the line wings is (Verhamme, Schaerer \& Maselli 2006)
\begin{equation}
\tau \sim 0.26 T^{-1/2}_4 N_{{\rm H},20} v^{-2}_{w,8} b_{12.85},
\end{equation}
where $T = 10^4 T_4$ K and $N_{\rm H} = N_{{\rm H},20} 10^{20}$ cm$^{-2}$ are the temperature and obscuring hydrogen column density of the medium, respectively.  The turbulent velocity in the interstellar medium is $b = 12.85 b_{12.85}$ km s$^{-1}$ (Verhamme, Schaerer \& Maselli 2006), while $v_w = 1000 v_{w,8}$ km s$^{-1}$ is the velocity of the emitting atom in the line wings.  Multiple scattering is important for $\tau \gtrsim 0.3$ (Chevalier 1986) and any realistic treatment of non-thermal Ly$\alpha$ lines in a young galaxy has to include radiative transfer effects, which we have neglected in our analysis.

\begin{acknowledgements}
K.H. is grateful to: Ratko Janev, C.D. Lin and Fernando Mart\'{i}n for invaluable advice
regarding atomic cross sections; Dick McCray, Roger Chevalier,
Rob Fesen, Bob Kirshner and Bryan Gaensler for engaging discussions;
Christian Balan\c{c}a for providing atomic cross sections in an
electronic form; John Raymond and Mike Shull for helpful suggestions following their careful reading
of the manuscript.  He acknowledges the Max Planck Institutes for
Astrophysics (MPA) and Extraterrestrial Physics (MPE) for their
generous support and kind hospitality during the months of June to
October 2007, where he was a visiting postdoctoral scientist.  He is
indebted to the tranquil Bavarian countryside for necessary moments of
academic solitude, and to his wife, Stefanie, for her steadfast support.
\end{acknowledgements}



\begin{table}
\begin{center}
\caption{H$\alpha$ \& Ly$\beta$ Observations of SNRs}
\label{table:obs}
\begin{tabular}{lcccccccc}
\hline\hline
\multicolumn{1}{c}{Object} & \multicolumn{1}{c}{$t_{\rm{age}}$} &
\multicolumn{1}{c}{$d$} & 
\multicolumn{1}{c}{$v_s^\spadesuit$} &
\multicolumn{1}{c}{$\Re_{bn}$} &
\multicolumn{1}{c}{H$\alpha^\dagger$} &
\multicolumn{1}{c}{Ly$\beta^\dagger$} & 
\multicolumn{1}{c}{SN Type} &
\multicolumn{1}{c}{References}\\
\multicolumn{1}{c}{} & \multicolumn{1}{c}{(yr)} & \multicolumn{1}{c}{(kpc)} &
\multicolumn{1}{c}{(km s$^{-1}$)} & \multicolumn{1}{c}{} & \multicolumn{1}{c}{(erg s$^{-1}$)} &
\multicolumn{1}{c}{(erg s$^{-1}$)} & \multicolumn{1}{c}{} & \multicolumn{1}{c}{}\\
\hline
0505---67.9 & $\sim 10,000$ & 50 & 464---744 & $\gtrsim 0.7$ & $7.15_{34}$ & $(4.58 \pm 0.23)_{34}$ & Ia & 5,7,10,12 \\
0509---67.5 & $\gtrsim 250$ & 50 & 5200---6300 & --- & $\lesssim 3.29_{34}$ & $\lesssim(2.33 \pm 0.18)_{34}$ & Ia & 5,7,9,10,12 \\
0519---69.0 & $ \gtrsim 400$ & 50 & 1032---1809 & $0.8 \pm 0.2$ & $3.19_{34}$ & $(2.93 \pm 0.07)_{34}$ & Ia & 5,7,9,10,12 \\
SN 1006 & 992 &  $2.0^{+0.4}_{-0.3}$ & 2290---3111 & $0.84^{+0.03}_{-0.01}$ & $(4.20 \pm 0.92)_{31}$  & --- & Ia & 4,7 \\
Kepler & 384 & $2.9 \pm 0.4$ & 1518---2446 & $0.72 \pm 0.37$ & $6.74_{31}$  & --- & --- & 1,7\\ 
RCW 86 & 1802$^\ddagger$ & 2.5 & 496---662 & $1.18 \pm 0.03$ & $(2.09 \pm 0.25)_{30}$ & --- & --- & 3,7,9,13 \\
SNR 1987A & 18 & 50 & 7840---9200 & 1? & $(2.98 \pm 0.33)_{34}$ & --- & II & 6,7,11 \\
Tycho & 406 & 1.5---3.1 & 1631---2344 & $0.67 \pm 0.1$ & $(4.15 \pm 2.31)_{30}$ & --- & Ia & 2,3,7,10  \\  
\hline
\end{tabular}
\end{center}
\scriptsize
$\dagger$ Inferred broad line luminosities from published line fluxes.  Note that $A_b$ is shorthand notation for $A \times 10^b$.\\
$\ddagger$ Assuming that RCW 86 is the remnant of SN 185.\\
$\spadesuit$ Quoted for $0.1 \le \beta \le 1$, where $\beta = T_e/T_p$ is the ratio of electron to proton temperature.\\
1: Blair, Long \& Vancura (1991);
2: Chevalier, Kirshner \& Raymond (1980);
3: Ghavamian et al. (2001);
4: Ghavamian et al. (2002);
5: Ghavamian et al. (2007);
6: Heng et al. (2006);
7: Heng \& McCray (2007);
8: Long \& Blair (1990);
9: Rest et al. (2005);
10: Smith et al. (1991);
11: Smith et al. (2005);
12: Tuohy et al. (1982);
13: Zombeck (1982).
\normalsize
\end{table}

\begin{table}
\begin{center}
\caption{Predicted broad Ly$\alpha$, Ly$\beta$, Ly$\gamma$, H$\beta$, P$\alpha$ and broad/narrow two-photon luminosities (erg s$^{-1}$)}
\label{table:predict}
\begin{tabular}{lccccccc}
\hline\hline
\multicolumn{1}{c}{Object} &
\multicolumn{1}{c}{Ly$\alpha$} &
\multicolumn{1}{c}{Ly$\beta$} &
\multicolumn{1}{c}{Ly$\gamma$ }&
\multicolumn{1}{c}{H$\beta$} &
\multicolumn{1}{c}{P$\alpha$} &
\multicolumn{1}{c}{Broad 2$\gamma$} &
\multicolumn{1}{c}{Narrow 2$\gamma$}\\
\hline
0505---67.9 & $(4.90 \pm 4.61)_{36}$ & $(1.09 \pm 0.34)_{36}$ & --- &
--- & --- & $(6.14 \pm 1.38)_{35}$ & $(2.51 \pm 0.12)_{35}$\\
0509---67.5 & $(2.68 \pm 2.05)_{35}$ & $(1.29 \pm 0.14)_{35}$ & $(2.04
\pm 0.08)_{34}$ & $(5.97 \pm 0.36)_{33}$ & $(1.07 \pm 0.06)_{33}$ &
$(6.16 \pm 0.02)_{35}$ & ---\\
0519---69.0 & $(1.46 \pm 1.29)_{36}$ & $(2.93 \pm 1.45)_{35}$ & --- &
--- & --- & $(5.00 \pm 1.37)_{35}$ & $(2.86 \pm 1.83)_{35}$\\
SN 1006 & $(8.42 \pm 2.57)_{32}$ & $(2.54 \pm 0.15)_{32}$ & --- & ---
& --- & $(5.91 \pm 2.04)_{32}$ & $(2.97 \pm 2.03)_{32}$\\
Kepler & $(2.95 \pm 1.24)_{33}$ & $(5.30 \pm 2.18)_{32}$ & --- & --- &
--- & $(1.01 \pm 0.15)_{33}$ & $(5.82 \pm 3.70)_{32}$\\ 
RCW 86 & $(2.30 \pm 0.94)_{32}$ & $(3.48 \pm 1.47)_{31}$ & --- & --- &
--- & $(1.76 \pm 0.48)_{31}$ & $(4.11 \pm 0.06)_{30}$\\
SNR 1987A & $(2.96 \pm 0.05)_{35}$ & $(7.64 \pm 0.59)_{34}$ & $(1.99
\pm 0.44)_{34}$ & $(5.96 \pm 1.40)_{33}$ & $(1.06 \pm 0.25)_{33}$ &
$(6.43 \pm 0.52)_{35}$ & $(5.75 \pm 0.70)_{35}$\\
Tycho & $(1.03 \pm 0.04)_{32}$ & $(2.32 \pm 0.44)_{31}$ & --- & --- &
--- & $(6.83 \pm 4.52)_{31}$ & $(5.38 \pm 4.78)_{31}$\\  
\hline
\end{tabular}
\end{center}
\scriptsize
Note: $A_b$ is shorthand notation for $A \times 10^b$.
\normalsize
\end{table}

\begin{table}
\begin{center}
\caption{Radiative decay fractions}
\label{table:einstein}
\begin{tabular}{lcc}
\hline\hline
\multicolumn{1}{c}{Quantity} &
\multicolumn{1}{c}{Value} &
\multicolumn{1}{c}{Relevance}\\
\hline
$B_{3p,2s}$ & 0.1183 & H$\alpha$, Ly$\beta$\\
$B_{4s,2p}$ & 0.5841 & H$\beta$, P$\alpha$\\
$B_{4s,3p}$ & 0.4159 & H$\beta$ P$\alpha$\\
$B_{4p,1s}$ & 0.8402 & H$\beta$, Ly$\gamma$, P$\alpha$\\
$B_{4p,2s}$ & 0.1191 & H$\beta$, Ly$\gamma$, P$\alpha$\\
$B_{4p,3s}$ & 3.643$_{-2}$ & H$\beta$, Ly$\gamma$, P$\alpha$\\
$B_{4p,3d}$ & 4.282$_{-3}$ & H$\beta$, Ly$\gamma$, P$\alpha$\\
$B_{4d,2p}$ & 0.7456 & H$\beta$, P$\alpha$\\
$B_{4d,3p}$ & 0.2544 & H$\beta$, P$\alpha$\\
\hline
\end{tabular}
\end{center}
\scriptsize
Note that $A_b$ is shorthand notation for $A \times 10^b$.
\normalsize
\end{table}

\appendix

\section{Ratio of Einstein A-coefficients}
\label{append:einstein}

To compute the rate coefficients for Ly$\alpha$, Ly$\beta$, Ly$\gamma$, H$\alpha$, H$\beta$ and P$\alpha$, one needs to calculate the ratio of Einstein A-coefficients.  The Einstein A-coefficient for hydrogen, $A_{nl,n^\prime l^\prime}$, is the radiative decay rate (s$^{-1}$) from the levels $nl$ to $n^\prime l^\prime$.  The radiative decay fraction is
\begin{equation}
B_{nl,n^\prime l^\prime} = A_{nl,n^\prime l^\prime} \left (\sum_{n^{\prime \prime} l^{\prime \prime}} A_{nl,n^{\prime \prime} l^{\prime \prime}} \right)^{-1},
\end{equation}
where the sum is over all transitions $nl \rightarrow n^{\prime \prime} l^{\prime \prime}$ permitted by the electric dipole selection rule, $l^{\prime \prime} = l \pm 1$.  A list of the relevant radiative decay fractions is listed in Table \ref{table:einstein}.  For example, to compute $\Gamma_{\rm{H}\beta/\rm{H}\alpha}(nl,n^\prime l^\prime)$, we need to consider $nl = B_{4s,2p} 4s + B_{4p,2s} 4p + B_{4d,2s} 4d$.  Our computed value of $B_{3p,2s} = 0.1183$ is in close agreement with the 0.1184 value quoted by Mart\'{i}n (1999).

The Einstein A-coefficients are proportional to the square of the
magnitude of the radial integrals, $\vert R^{n^\prime l^\prime}_{nl}
\vert^2$.  (See Appendix A2 of HM07 for details on how to calculate them
analytically.)  As a check, we have compared our computed values of $\vert R^{n^\prime l^\prime}_{nl} \vert^2$ to the ones tabulated by Green, Rush and Chandler (1957), and find them to be in agreement.

\section{Atomic Cross Sections}
\label{append:atomic}

\begin{figure}
\resizebox{\hsize}{!}{\includegraphics{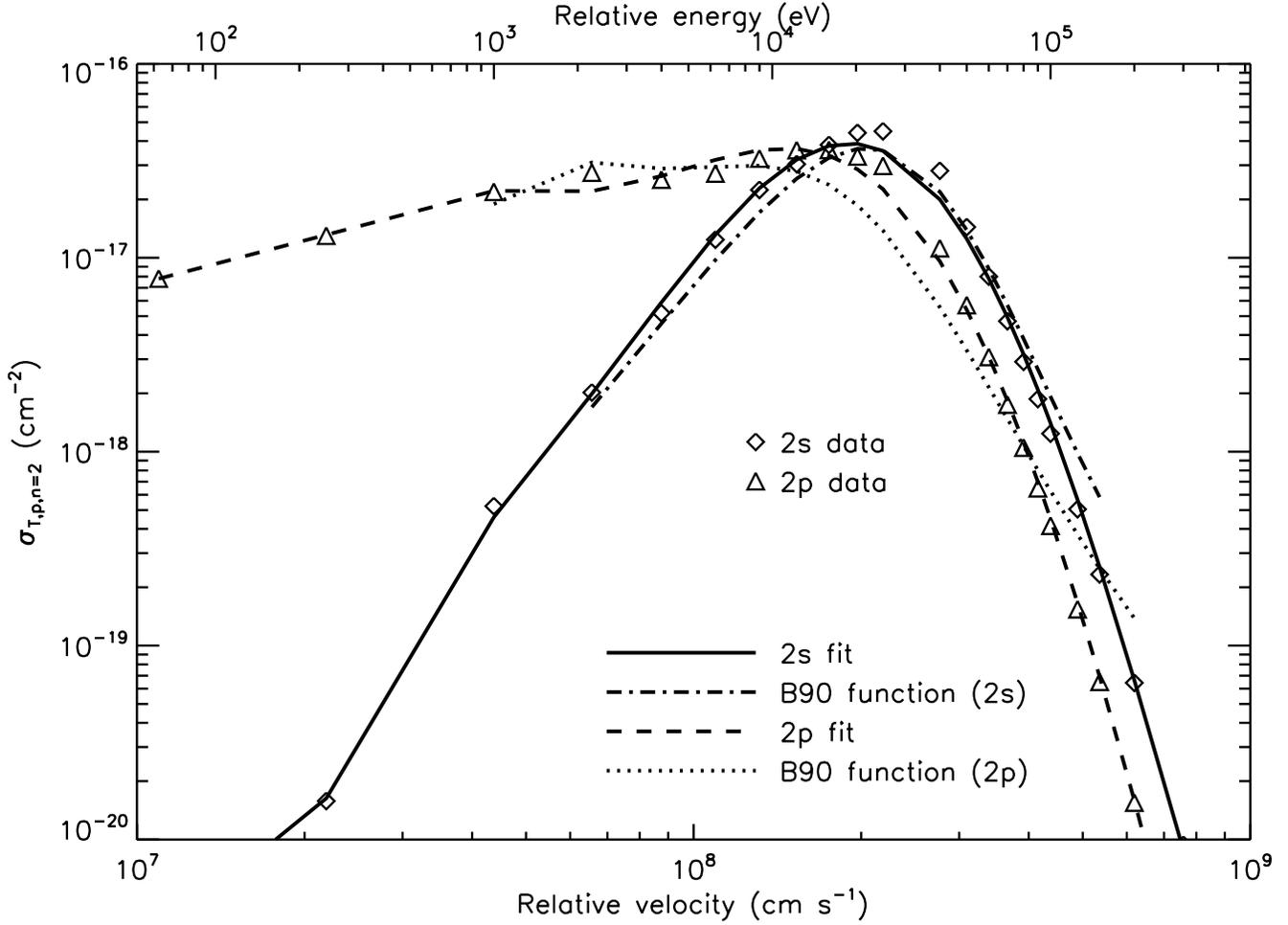}}
\caption{Cross sections for charge transfers between hydrogen atoms
and protons, to the sub-levels $2s$ and $2p$, taken from Belk\'{i}c, Gayet
\& Salin (1992) and Harel, Jouin \& Pons (1998).  Also shown are the fitting functions of Barnett et al. (1990), denoted by ``B90''.}
\label{fig:pct2}
\end{figure}

\begin{figure}
\resizebox{\hsize}{!}{\includegraphics{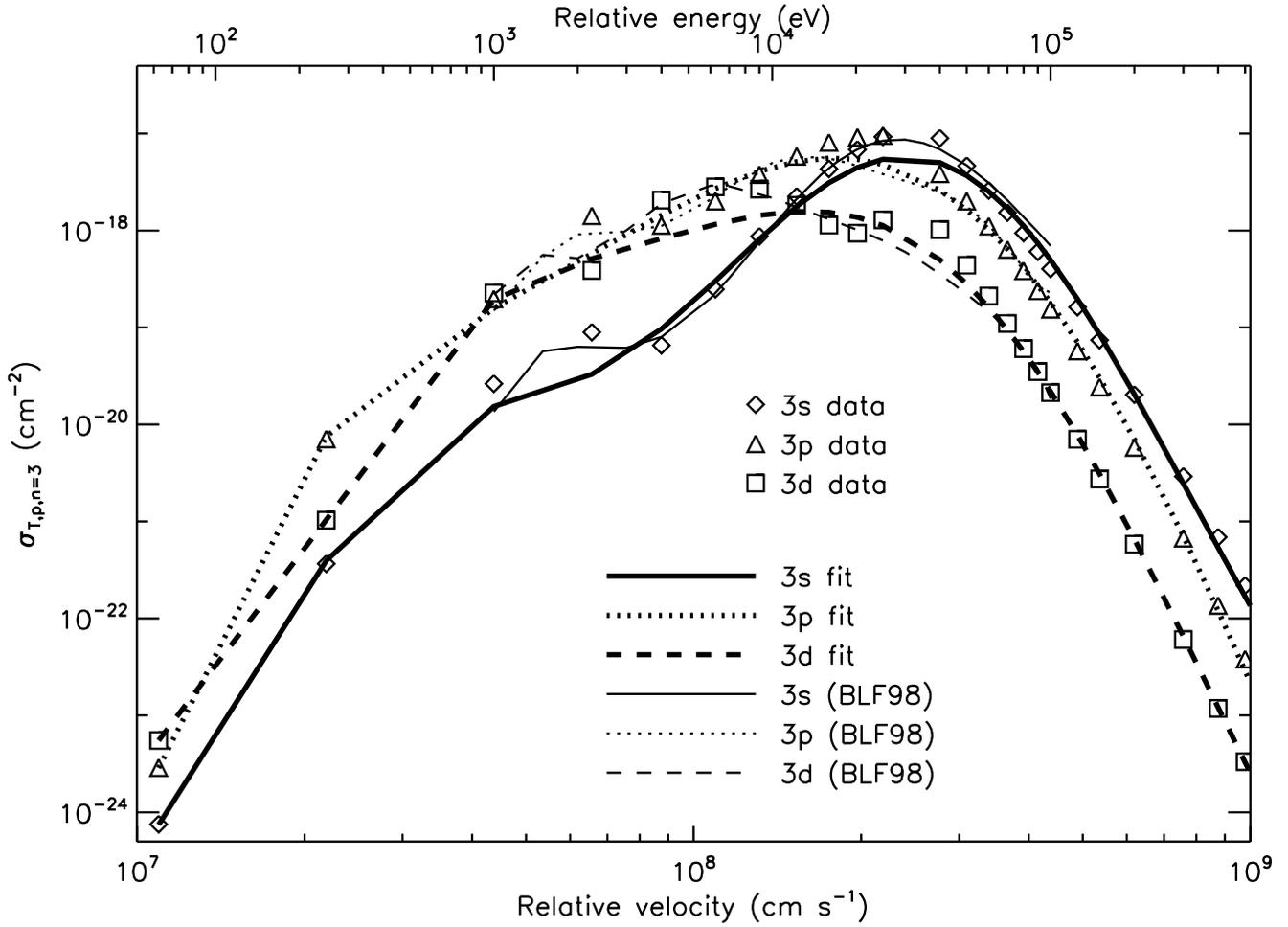}}
\caption{Cross sections for charge transfers between hydrogen atoms
and protons, to the sub-levels $3s$, $3p$ and $3d$, taken from
Belk\'{i}c, Gayet \& Salin (1992) and Harel, Jouin \& Pons (1998).
Shown for comparison are the calculations of Balan\c{c}a, Lin \&
Feautrier (1998), denoted by ``BLF98''.}
\label{fig:pct3}
\end{figure}

\begin{figure}
\resizebox{\hsize}{!}{\includegraphics{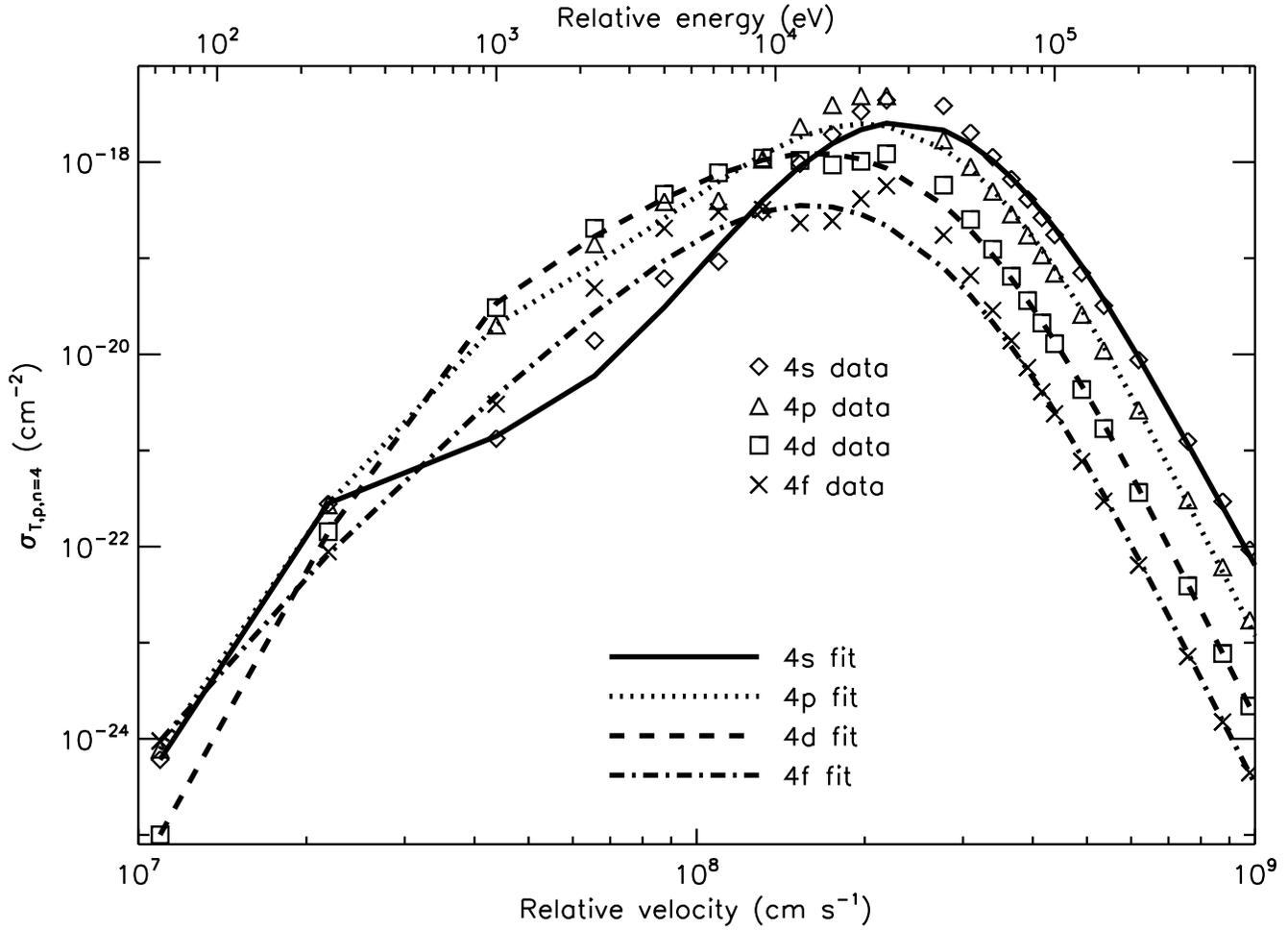}}
\caption{Cross sections for charge transfers between hydrogen atoms
and protons, to the sub-levels $4s$, $4p$, $4d$ and $4f$, taken from
Belk\'{i}c, Gayet \& Salin (1992) and Harel, Jouin \& Pons (1998).}
\label{fig:pct4}
\end{figure}

\begin{figure}
\resizebox{\hsize}{!}{\includegraphics{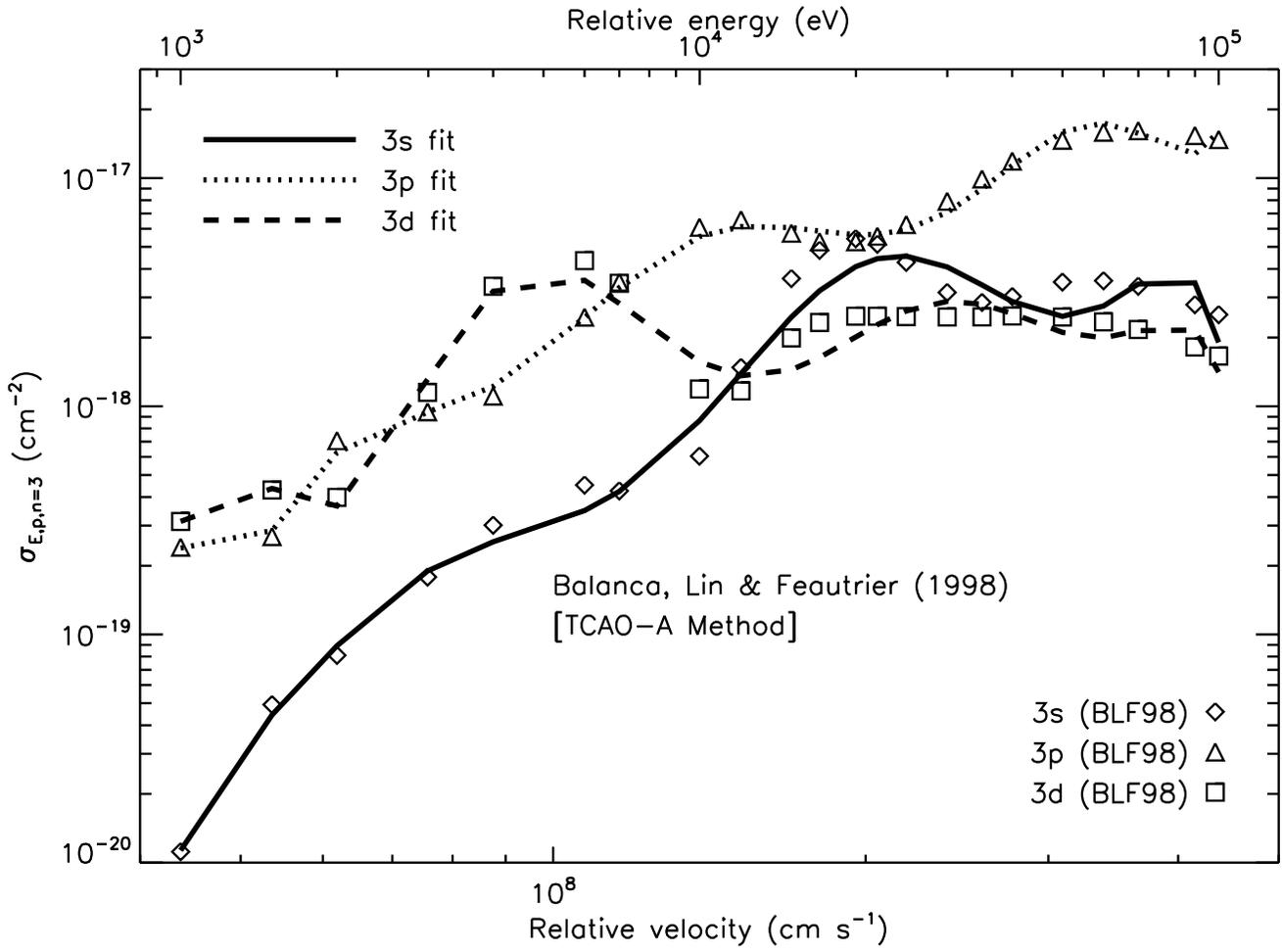}}
\caption{Cross sections for impact excitation of hydrogen atoms by
protons, from the two-center atomic-orbital (TCAO) close-coupling
calculations of Balan\c{c}a, Lin \& Feautrier (1998).  Shown are the
fits to the TCAO-A calculations.}
\label{fig:pexc3}
\end{figure}

\begin{figure}
\resizebox{\hsize}{!}{\includegraphics{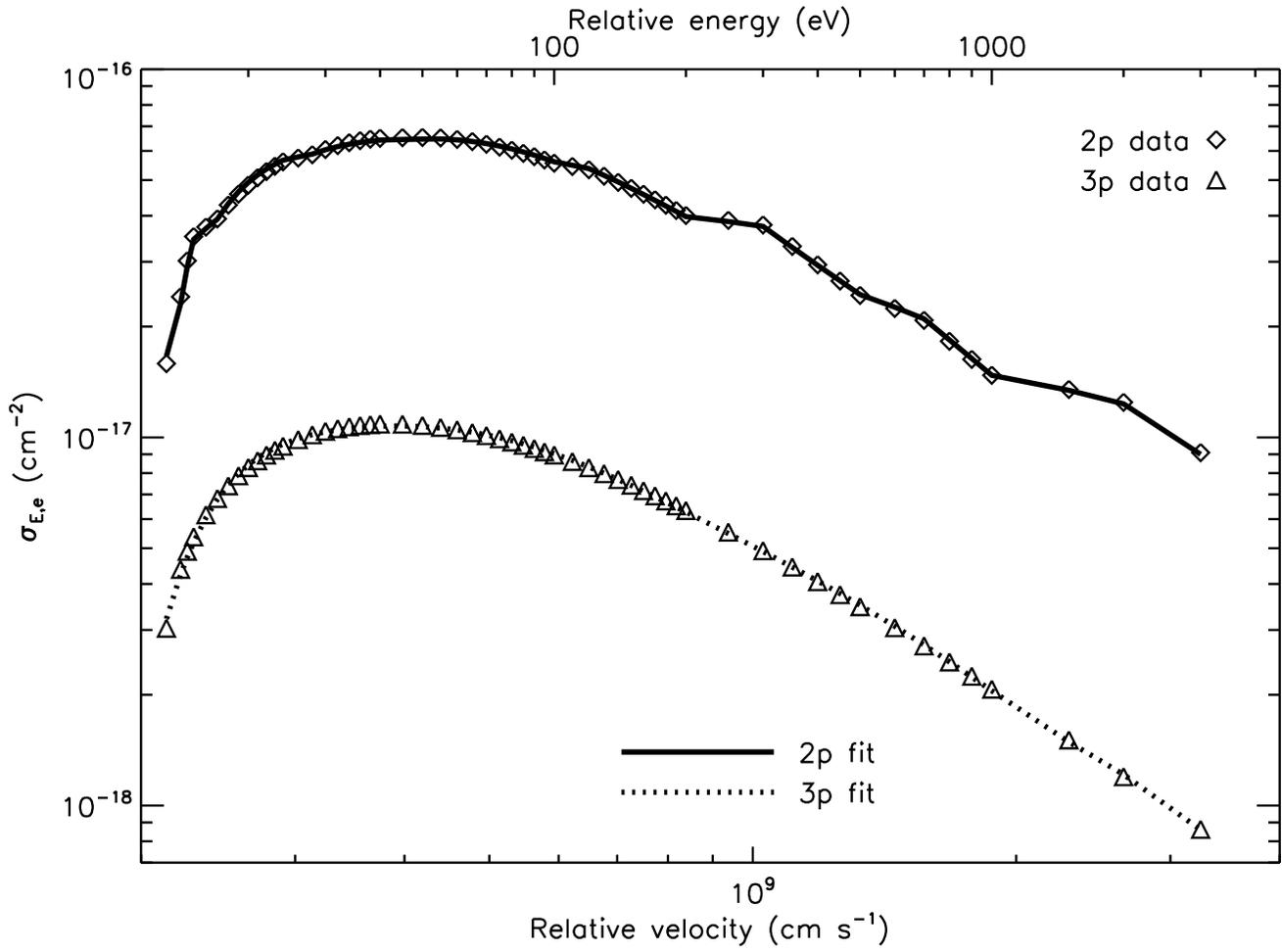}}
\caption{Cross sections for impact excitation of hydrogen atoms by electrons, to the sub-levels $2p$ and $3p$, from the {\it NIST Electron-Impact Cross Section Database}.}
\label{fig:eexc}
\end{figure}

Cross sections for interactions between protons and hydrogen atoms
(charge transfer and excitation) to the sub-levels $3s$, $3p$ and $3d$ are
computed in Balan\c{c}a, Lin \& Feautrier (1998) and kindly provided to us by
C. Balan\c{c}a (2007, private communication).  We note that these calculations utilize a two-center atomic-orbital (TCAO) close-coupling method with an asymmetric (TCAO-A) basis set of 26 states, so as to avoid spurious numerical oscillations
caused by using a traditional, symmetric set (TCAO-S).  We fit these cross
sections, as well as those from  Belk\'{i}c, Gayet \& Salin (1992),
Harel, Jouin \& Pons (1998) and the {\it NIST Electron-Impact Cross
Section Database} using the function:
\begin{equation}
{\cal F}\left(x;\vec{A}\right) = \exp{\left(\frac{A_0}{2} + \sum^{8}_{i=1} ~A_i ~{\cal C}_i\left(x\right)\right)},
\end{equation}
where the coefficients $\vec{A}=A_i$ for $0 \le i \le 8$ are the
fitting parameters.  The quantities ${\cal C}_i$ are the Chebyshev orthogonal polynomials:
\begin{eqnarray}
{\cal C}_1\left(x\right) = x,\\
{\cal C}_2\left(x\right) = 2x^2 - 1,\\
{\cal C}_3\left(x\right) = 4x^3 - 3x,\\
{\cal C}_4\left(x\right) = 8\left(x^4 - x^2 \right) + 1,\\
{\cal C}_5\left(x\right) = 16x^5 - 20x^3 + 5x,\\
{\cal C}_6\left(x\right) = 32x^6 - 48x^4 + 18x^2 - 1,\\
{\cal C}_7\left(x\right) = 64x^7 - 112x^5 + 56x^3 - 7x,\\
{\cal C}_8\left(x\right) = 128x^8 - 256x^6 + 160x^4 - 32x^2 + 1.\\
\end{eqnarray}
The fitting variable $x$ is defined as
\begin{equation}
x= \frac{ \ln{\left(E/E_{\rm{min}}\right)} - \ln{\left(E_{\rm{max}}/E\right)} }{\ln{\left(E_{\rm{max}}/E_{\rm{min}}\right)}},
\end{equation}
where $E$ is the relative energy between the collidants;
$E_{\rm{min}}$ and $E_{\rm{max}}$ are the respective minimum and maximum energies
to which the data are available.  We use the Levenberg-Marquardt
algorithm, which combines the steepest descent and inverse-Hessian
function fitting methods, as implemented in {\tt IDL}.  The fits are
sensitive to the initial values of the parameters fed to the
algorithm; we use the values of the fit parameters for
$\sigma_{\rm{T,p},2s}$ in Barnett et al. (1990) as a guide.  In
providing ``measurement errors'' to our fitting algorithm, we assume a
fiducial error of 10\%.  Selected cross sections are shown in
Figs. \ref{fig:pct2}, \ref{fig:pct3}, \ref{fig:pct4}, \ref{fig:pexc3}
and \ref{fig:eexc}, while the fitting coefficients are presented in Table \ref{tab:fits}.

\begin{table}
\begin{center}
\caption{Fitting Parameters for Various Cross Sections}
\label{tab:fits}
\begin{tabular}{lccccccccc}
\hline\hline
\multicolumn{1}{c}{Reaction} & \multicolumn{1}{c}{$A_0$} & \multicolumn{1}{c}{$A_1$} & \multicolumn{1}{c}{$A_2$} & \multicolumn{1}{c}{$A_3$} & \multicolumn{1}{c}{$A_4$} & \multicolumn{1}{c}{$A_5$} & \multicolumn{1}{c}{$A_6$} & \multicolumn{1}{c}{$A_7$} & \multicolumn{1}{c}{$A_8$}\\
\hline
$\sigma_{\rm{T,p},2s}$ & $-9.76244_1$ & -$8.02645$ & $-9.03528$ & $-6.82148_{-1}$ & 1.41138 & $-8.43613_{-3}$ & $-1.31162_{-1}$ & $-3.07135_{-1}$ & $3.60166_{-1}$\\
$\sigma_{\rm{T,p},2p}$ & $-9.47856_1$ & $-1.36982_1$ & $-7.48411$ & -$6.55633_{-1}$ & 1.08973 & $3.45255_{-1}$ & $-2.02178_{-1}$ & $-2.99212_{-1}$ & $2.88704_{-1}$ \\
$\sigma_{\rm{T,p},3s}$ & $-1.03831_{2}$ & $-5.12682$ & $-1.03311_{1}$ & $-2.36941_{-1}$ & $7.99974_{-1}$ & $6.55927_{-1}$ & $-2.64604_{-1}$ & $-6.82557_{-1}$ & $7.65819_{-1}$ \\
$\sigma_{\rm{T,p},3p}$ & $-1.04117_{2}$ & $-9.31370$ & $-1.09011_{1}$ & $2.72437_{-1}$ & $8.33272_{-1}$ & $9.72800_{-1}$ & $-3.40175_{-1}$ & $-8.21497_{-2}$ & $1.06280_{-1}$ \\
$\sigma_{\rm{T,p},3d}$ & $-1.06187_{2}$ & $-8.82542$ & $-1.05048_{1}$ & $9.29637_{-1}$ & $9.65598_{-1}$ & $7.99638_{-3}$ & $1.21539_{-1}$ & $-6.52001_{-1}$ & $4.15571_{-1}$ \\
$\sigma_{\rm{T,p},4s}$ & $-1.05018_{2}$ & $-5.69159$ & $-9.63683$ & $-8.65652_{-1}$ & 1.06090 & $9.33260_{-1}$ & $-8.64227_{-1}$ & $-1.30903_{-1}$ & $4.32513_{-1}$ \\
$\sigma_{\rm{T,p},4p}$ & $-1.06245_{2}$ & $-7.18909$ & $-1.07639_{1}$ & $1.84045_{-1}$ & 1.10514 & $3.64508_{-1}$ & $-2.30906_{-1}$ & $-4.14293_{-1}$ & $4.42254_{-1}$ \\
$\sigma_{\rm{T,p},4d}$ & $-1.08456_{2}$ & $-7.54281$ & $-1.11995_{1}$ & 1.15545 & $9.55810_{-1}$ & $9.05249_{-2}$ & $-1.37453_{-1}$ & $-4.09046_{-1}$ & $3.40271_{-1}$ \\
$\sigma_{\rm{T,p},4f}$ & $-1.10117_{2}$ & $-8.87472$ & $-1.00318_{1}$ & $3.41036_{-1}$ & 1.43444 & $1.22887_{-2}$ & $-4.35737_{-1}$ & $-7.29281_{-2}$ & $1.87428_{-1}$ \\
$\sigma_{\rm{E,e},2p}$ & $-7.67030_{1}$ & $-9.47539_{-1}$ & $-6.80475_{-1}$ & $2.77805_{-1}$ & $-1.18599_{-1}$ & $6.65029_{-2}$ & $-4.38768_{-2}$ & $2.20102_{-2}$ & $-2.24010_{-2}$ \\
$\sigma_{\rm{E,e},3p}$ & $-8.02681_{1}$ & $-1.00579$ & $-6.24421_{-1}$ & $2.59918_{-1}$ & $-1.13262_{-1}$ & $6.48909_{-2}$ & $-4.34274_{-2}$ & $2.15313_{-2}$ & $-2.31916_{-2}$ \\
$\sigma_{\rm{E,p},3s}$ & $-8.43473_{1}$ & 2.69180 & $-8.94871_{-1}$ & $-1.25347_{-1}$ & $-1.27427_{-1}$ & $2.16335_{-1}$ & $7.34001_{-3}$ & $-2.19545_{-1}$ & $-1.76763_{-1}$ \\
$\sigma_{\rm{E,p},3p}$ & $-8.10541_{1}$ & 2.27478 & $-4.95559_{-1}$ & $1.56523_{-2}$ & $2.45763_{-2}$ & $-1.29970_{-1}$ & $-2.66319_{-2}$ & $-6.13906_{-2}$ & $2.46815_{-1}$ \\
$\sigma_{\rm{E,p},3d}$ & $-8.22914_{1}$ & $7.43945_{-1}$ & $-5.04030_{-1}$ & $5.76290_{-2}$ & $8.40403_{-2}$ & $-3.04995_{-1}$ & $7.62616_{-2}$ & $2.56426_{-1}$ & $-3.65589_{-1}$ \\
\hline
\end{tabular}
\end{center}
\scriptsize
Note: $A_b$ is shorthand for $A \times 10^b$.  The following subscripts are used: ``T'' (charge transfer), ``E'' (impact excitation), ``p'' (proton) and ``e'' (electron).\\
\normalsize
\end{table}

\section{Broad Emission from Within the Shock Front?}
\label{append:within}

We have considered only the case of hydrogen atoms crossing the shock front (in the frame of the front) and interacting with protons.  In this sub-section, we examine the possibility of Ly$\alpha$ being created {\it within} the shock front.  The width of the shock front in collisional shocks is on the order of an atomic mean free path, $l_{\rm{mfp}}$, assuming a pure hydrogen gas.  Zel'dovich \& Raizer (1966) have shown that for weak shocks, the collisional shock width is
\begin{equation}
\delta \sim l_{\rm{mfp}} ~\frac{p_0}{p_1-p_0},
\end{equation}
where $p_0$ and $p_1$  are the pre- and post-shock fluid pressures, both in the case of a viscous shock with no heat conduction and for a heat-conducting shock with no viscosity.  When the change in pressure across the shock front is comparable to the magnitude of the pre-shock pressure, $\delta \sim l_{\rm{mfp}}$ (see also Landau \& Lifshitz 1963).  Even in the limit of infinite Mach number, Sakurai (1957) finds that $\delta/l_{\rm{mfp}} = 1.42$.

The question is: how robust is the assumption of shocks in Balmer-dominated SNRs being collisionless?  This occurs when the electron and proton gyroradii --- $r_e$ and $r_p$, respectively --- are much smaller than $l_{\rm{mfp}}$.  The typical value of the magnetic fields in SNRs is $B = B_{-4} 10^{-4}$ G.  For example, Strom \& Duin (1973) find $3 \times 10^{-4}$ and  $5 \times 10^{-4}$ G for Tycho and Cas A, respectively.  The electron gyroradius is
\begin{equation}
r_e \sim 10^5 \mbox{ cm} ~v_{e,8} B^{-1}_{-4},
\end{equation}
where $v_{e,8} = v/1000$ km s$^{-1}$ is the velocity of the electron.
For protons, we have $r_p \sim 10^8$ cm $v_{p,8}/B_{-4}$.  It follows
that the transition from collisionless to collisional shock occurs
when the gyroradii $\sim l_{\rm{mfp}}$, or when the density of
particles is $n_e \gtrsim 10^{10} v^{-1}_{e,8} \sigma^{-1}_{a,-15}
B_{-4}$ cm$^{-3}$ and $n_p \gtrsim 10^7 v^{-1}_{p,8}
\sigma^{-1}_{a,-15} B_{-4}$ cm$^{-3}$, where $\sigma_a =
\sigma_{a,-15} 10^{-15}$ cm$^{-2}$ is the typical value of the cross
section for atomic interactions (charge transfer and ionization; see
HM07 and H07).  These densities are much larger than typical values for the interstellar medium ($\sim 1$ cm$^{-3}$) or even for molecular clouds ($\sim 100$ to 1000 cm$^{-3}$).

About $\exp{(-l_s/l_{\rm{mfp}})}$ of the hydrogen atoms cross the shock front without being ionized, where $l_s$ is the width of the shock.  In collisional shocks, $l_s \sim \delta$.  In collisionless shocks, $l_s \sim r_e$, $l_s/l_{\rm{mfp}} \ll 1$ and virtually all of the atoms pass through.  We thus conclude that broad Ly$\alpha$ is probably not produced in a significant amount within the shock front, consistent with the findings of H07.

\end{document}